\documentclass[conference]{IEEEtran}
\IEEEoverridecommandlockouts

\usepackage{cite}
\usepackage{amsmath,amssymb}
\usepackage{booktabs}
\usepackage{graphicx}
\usepackage{url}
\usepackage{tikz}
\usetikzlibrary{positioning,fit,backgrounds,arrows.meta,calc}
\usepackage{hyperref}

\begin{document}

\title{World-Model-Aware Responsibility Allocation in\\
Heterogeneous Logistics Systems}

\author{
  
\IEEEauthorblockN{Artan Markaj\thanks{\copyright~2026 IEEE. Personal use of this material is permitted. Permission from IEEE must be obtained for all other uses, in any current or future media, including reprinting/republishing this material for advertising or promotional purposes, creating new collective works, for resale or redistribution to servers or lists, or reuse of any copyrighted component of this work in other works.}}
\IEEEauthorblockA{\textit{Eurogate GmbH \& Co. KGaA, KG} \\
Hamburg, Germany\\
artan.markaj@eurogate.eu}

\and

\IEEEauthorblockN{Niklas Jobs}
\IEEEauthorblockA{\textit{Helmut Schmidt University Hamburg} \\
Hamburg, Germany\\
niklas.jobs@hsu-hh.de}

\and

\IEEEauthorblockN{Felix Gehlhoff}
\IEEEauthorblockA{\textit{Helmut Schmidt University Hamburg} \\
Hamburg, Germany\\
felix.gehlhoff@hsu-hh.de}
}

\maketitle

\begin{abstract}
Logistics systems increasingly mix \emph{autonomous logistic equipment} (ALE)
with non-autonomous machinery under a central control system (CS), where the
best decision-maker depends on who holds the most current world model, yet
authority is fixed at design time. When an ALE's local model and the CS global
model diverge, both act on incompatible beliefs and produce deadlocks that
resource-based handling neither explains nor prevents. We propose the
World-Model-Aware Responsibility Framework (WMARF), which assigns authority
dynamically from CS world-model quality and equipment automation level, and
classifies deadlocks by the state of authority -- none, in transition, or
divergent. In a discrete-event simulation of two ALE converging on a
semi-automated transfer point, reproduced over the VDA~5050 interface, a
divergence deadlock under static control is prevented by a proximity-triggered
handoff. Because authority follows information quality rather than a shared
protocol, the scheme stays valid as autonomy grows.
\end{abstract}


\section{Introduction}
Logistics systems increasingly combine equipment at very different levels of
automation. \emph{Autonomous logistic equipment} (ALE) (e.g., automated guided
vehicles, autonomous mobile robots, and increasingly autonomous
trucks) operates alongside non- or semi-autonomous equipment such as
conveyors, handling stations, and manually operated machinery, all coordinated
by a central control system
(CS)~\cite{fragapane2021planning,keith2024amrreview}. The CS maintains a
global world model and issues coordination decisions, while autonomous
equipment carries a rich, real-time \emph{local} world model from onboard
sensors; non-autonomous equipment, by contrast, contributes only coarse state
information whose granularity falls with its automation level.

This heterogeneity creates a structural problem that current control
architectures do not address: \emph{which entity should hold decision
authority is not fixed, it depends on who has the most complete and current
world model for the decision at hand}. Yet responsibility is almost always
allocated statically, at design time, from the automation landscape as it
stands then~\cite{gehlhoff2025interaction}, not from the information quality
actually available at runtime. When the local model of an ALE
diverges from the CS model (through latency, sensor-coverage gaps, or
equipment that does not report fine-grained state), neither entity has clear
authority, and both may act on locally consistent but globally incompatible
beliefs~\cite{gehlhoff2018imperfect}, proceeding at once into a deadlock that
no party flagged.

Two forces make this acute. First, a \emph{maturity trap}: autonomous
equipment will grow markedly more capable within a system's operational life,
so a centralized architecture chosen today for safety can become tomorrow's
bottleneck, while a decentralized one chosen for future scalability may be
unsafe now (current ALE may lack the sensing or reasoning maturity to act
safely without CS oversight). Second, a \emph{multi-vendor} reality: equipment from different
manufacturers exposes incompatible interfaces and state models, so neither
full centralization (one common protocol) nor full decentralization (every
device negotiating autonomously) is achievable.

We address this with the World-Model-Aware Responsibility Framework (WMARF), a
reference architecture for runtime authority assignment. It raises two
questions: whether decision authority between autonomous equipment and the CS
can be assigned \emph{dynamically} from world-model quality and automation
level (RQ1), and whether the resulting deadlocks can be classified by the state
of authority so that each admits a targeted resolution (RQ2). Accordingly we
make two contributions: (i)~WMARF, a world-model-aware scheme that assigns
authority dynamically and labels the resulting deadlocks by their authority
state -- none, in transition, or divergent; and (ii)~the framing of the
\emph{maturity trap} as a design problem, with a migration path that keeps the
allocation valid as equipment autonomy grows without architectural redesign,
and without requiring vendors to agree on a protocol.

\section{Background and Related Work}
Coordination of equipment in logistics has matured along two poles. At
one, a central control system (CS) schedules, routes, and dispatches from a
global model~\cite{fragapane2021planning,gehlhoff2025interaction}, in the
tradition of human supervisory control~\cite{sheridan1992supervisory}; at the
other, ALE negotiates locally and decentralizes decisions to react to a
changing environment~\cite{keith2024amrreview,meseguer2025taskalloc}.
Decentralized, agent-based schemes increasingly reason about imperfect
information~\cite{gehlhoff2018imperfect}, yet still assume the deciding
entities are capable agents.

Reassigning authority to whoever is best placed to decide is the subject of
\emph{variable autonomy}~\cite{theodorou2024variable} -- a topic spanning
adjustable and sliding autonomy~\cite{sellner2006sliding}, mixed-initiative
interaction~\cite{horvitz1999mixed}, and dynamic role and task allocation for
human--robot teams~\cite{lamon2023roles,yuan2025adaptive}, grounded in classic
models of automation levels and function
allocation~\cite{parasuraman2000model}, and recently extended to \emph{learn}
when to adjust the level of automation online~\cite{hajnorouzi2025loa}. This
body almost exclusively arbitrates between a \emph{human} and an autonomous
system, keying the hand-over on human workload, trust, or task demand. Two of
its assumptions break in our setting: authority shifts between \emph{software systems}
(the CS and the ALE), not between a human and a software system; and one party -- non-autonomous
equipment that only reports coarse state -- is not a capable agent at all.
Deadlocks, in turn, are treated as resource or routing conflicts, from
classical AGV conflict resolution~\cite{vis2006survey,reveliotis2000conflict}
to recent learning- and barrier-based
avoidance~\cite{mueller2025marl,zhang2025deadlock}, and are never classified
by \emph{who should have been deciding}.

This leaves a gap that real installations occupy. Much work naturally
concentrates on the two autonomous extremes, fully centralized or fully
decentralized, while many deployed systems operate in the \emph{middle
ground}: mixed fleets from several vendors, ALE sharing space with conveyors,
cranes, and manual stations whose state the CS barely observes. Heterogeneous
multi-robot systems are themselves well surveyed~\cite{rizk2019heterogeneous},
but they coordinate \emph{robots}; here the heterogeneity spans autonomous and
non-autonomous equipment under one CS. Three failure patterns characterize it. A \emph{governance vacuum} arises when local and
global models diverge and no protocol says who resolves it.
\emph{Automation-level blindness} arises when the CS represents all equipment
uniformly, so decisions near poorly observable equipment are systematically
overconfident. And many deadlocks are \emph{governance failures}: two entities
act at once because authority was never transferred, not because a resource is
busy. WMARF targets this heterogeneous middle ground.

\section{The WMARF Framework}
\begin{figure}[t]
  \centering
\resizebox{\columnwidth}{!}{%
\begin{tikzpicture}[font=\footnotesize]
  \def\w{3.7}\def\h{2.0}
  \draw[thick] (0,0) rectangle (2*\w,2*\h);
  \draw[thick] (\w,0) -- (\w,2*\h);
  \draw[thick] (0,\h) -- (2*\w,\h);

  \node[align=center, text width=3.3cm, fill=black!4] at (0.5*\w,1.5*\h)
    {\textbf{Q2 -- Shared / Handoff}\\[2pt]\scriptsize both valid but may\\diverge; explicit protocol};
  \node[align=center, text width=3.3cm, fill=blue!5] at (1.5*\w,1.5*\h)
    {\textbf{Q1 -- CS-dominant}\\[2pt]\scriptsize CS model reliable;\\ALE follows CS};
  \node[align=center, text width=3.3cm, fill=red!7] at (0.5*\w,0.5*\h)
    {\textbf{Q4 -- Contested}\\[2pt]\scriptsize no reliable model;\\halt + escalate};
  \node[align=center, text width=3.3cm, fill=black!4] at (1.5*\w,0.5*\h)
    {\textbf{Q3 -- ALE-dominant}\\[2pt]\scriptsize onboard sensors are\\ground truth};

  \node[rotate=90, anchor=south, font=\bfseries] at (-0.65,\h) {CS world-model quality};
  \node[anchor=east, font=\scriptsize] at (-0.12,1.5*\h) {High};
  \node[anchor=east, font=\scriptsize] at (-0.12,0.5*\h) {Low};

  \node[anchor=north, font=\bfseries] at (\w,-0.6) {Equipment automation level};
  \node[anchor=north, font=\scriptsize] at (0.5*\w,-0.12) {Low};
  \node[anchor=north, font=\scriptsize] at (1.5*\w,-0.12) {High};
\end{tikzpicture}%
}
  \caption{The WMARF responsibility quadrants. Authority follows the entity
  with the better world model; an interaction's position shifts at runtime
  with proximity, model staleness, and sensor-vs-CS divergence.}
  \label{fig:concept}
\end{figure}
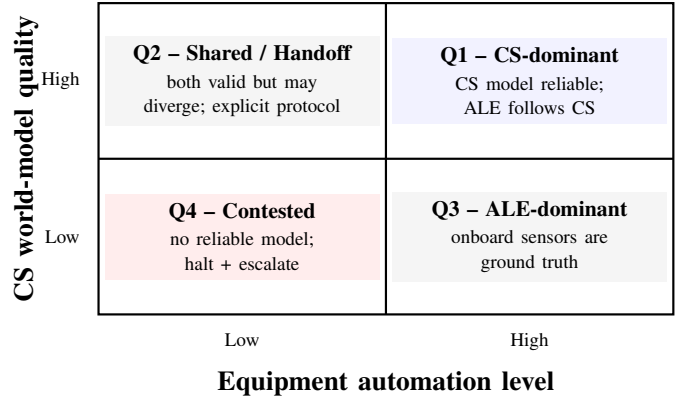

WMARF assigns decision authority dynamically by positioning each interaction
between an ALE and a target equipment in a two-dimensional space
(Fig.~\ref{fig:concept}). The \emph{structural} axis is
the target's automation level and, with it, how observable its state is to the
CS through its interface. Automation and observability usually rise together,
but not always: a highly automated device from another vendor may still expose
only coarse state and therefore sit lower on this axis than its autonomy alone
would suggest. The \emph{dynamic} axis is the current quality -- recency and
consistency -- of the CS world model for that target. Reading the structural
axis as observability is what lets WMARF absorb the multi-vendor problem:
authority follows how well a target can be observed, not which vendor built it
or whether a common protocol exists.

\subsection{Responsibility quadrants}
The two axes yield four authority zones. In \textbf{Q1} (high quality, high
automation) the CS model is reliable and the ALE follows CS instructions.
In \textbf{Q2} (high quality, low automation) the CS has a good global picture
but the target's internal state is under-represented; a handoff (transfer
of decision authority) gives the ALE local authority for the final approach, after which it reports back
and the CS updates its model. In \textbf{Q3} (low quality, high automation)
the CS model is stale or inconsistent -- post-fault, during reconfiguration, or
when update latency exceeds an age $\tau$ -- and the ALE trusts its onboard
sensors under a mandatory escalation flag. In \textbf{Q4} (low quality, low
automation) neither entity has a reliable model and the conservative
default -- halt, escalate, await clearance -- applies. Position is not static:
it shifts at runtime with ALE proximity, model staleness, and the
discrepancy between sensor readings and CS belief.

\subsection{Handoff triggers}
Authority transfers fire on measurable conditions. Table~\ref{tab:triggers}
lists an \emph{exemplary}, non-exhaustive set: CS$\rightarrow$ALE triggers on
proximity, model divergence, target automation class, or CS staleness, and
ALE$\rightarrow$CS triggers on completion or a locally detected deadlock; each
transfer is closed by an acknowledgment handshake. Identifying the full trigger
set and calibrating its thresholds is where the principal work of this
framework lies, and is the focus of future work.

\begin{table}[t]
\caption{Exemplary handoff triggers (non-exhaustive).}
\label{tab:triggers}
\centering
\footnotesize
\begin{tabular}{@{}lll@{}}
\toprule
Trigger & Direction & Condition \\
\midrule
Proximity           & CS$\to$ALE & distance $<\theta$ \\
Divergence          & CS$\to$ALE & $|\text{sensor}-\text{CS}| > \delta$ \\
Automation level    & CS$\to$ALE & target manual/semi \\
Staleness           & CS$\to$ALE & CS update age $>\tau$ \\
Resolution complete & ALE$\to$CS & interaction finished \\
Escalation          & ALE$\to$CS & ALE detects deadlock \\
\bottomrule
\end{tabular}
\end{table}

\subsection{Deadlock taxonomy by authority state}
We classify deadlocks by the \emph{state of decision authority} rather than
the contested resource. \textbf{Type~A (no authority assigned):} both entities
act on their own beliefs with no handoff engaged, a governance vacuum,
typical of Q4; resolved by the conservative default firing before either
commits. \textbf{Type~B (authority in transition):} authority is mid-transfer and the receiver has not yet acknowledged or
updated its model (e.g., the CS has issued a handoff token but the ALE has
not yet committed to the new authority), leaving a decision vacuum in the
transfer window; resolved by the acknowledgment handshake. \textbf{Type~C (authority on divergent models):} authority is held,
but the holders act on locally consistent yet globally incompatible models;
resolved by model reconciliation, with sensor data taking precedence within
proximity $\theta$ of the target. The quadrants indicate \emph{where} each type tends to
arise, not its definition (Table~\ref{tab:deadlock}).

\begin{table}[t]
\caption{Deadlock taxonomy by authority state.}
\label{tab:deadlock}
\centering
\footnotesize
\begin{tabular}{@{}cllp{2.6cm}@{}}
\toprule
Type & Authority & Arises in & Resolution \\
\midrule
A & none assigned    & Q4    & halt + escalate \\
B & in transition    & any handoff & acknowledgment handshake \\
C & divergent models & Q3 / Q2 & model reconciliation (sensor precedence within $\theta$) \\
\bottomrule
\end{tabular}
\end{table}

\subsection{Maturity migration}
As equipment matures, the thresholds $\theta,\delta,\tau$ can be retuned to
enlarge the ALE-dominant zone without changing the architecture; in the
limits WMARF degrades gracefully toward full centralization (an omniscient CS)
or full decentralization (fully autonomous equipment). This is a possible answer to
the maturity trap, and, because authority follows information quality rather
than a shared protocol, to the multi-vendor problem.

\section{Architecture and Implementation}
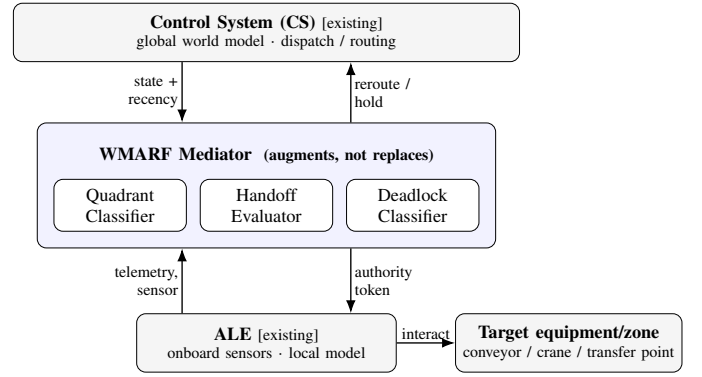
\begin{figure}[t]
  \centering
\resizebox{\columnwidth}{!}{%
\begin{tikzpicture}[
  font=\footnotesize,
  layer/.style={draw, rounded corners, align=center, minimum height=9mm, fill=black!4},
  mod/.style={draw, rounded corners, align=center, fill=white, text width=19mm,
              minimum height=8mm, inner sep=2pt},
  arr/.style={-{Latex[length=2mm]}, semithick},
  lbl/.style={font=\scriptsize, inner sep=1.5pt},
]
  \node[layer, minimum width=78mm] (cs)
    {\textbf{Control System (CS)} \scriptsize[existing]\\[-1pt]
     \scriptsize global world model $\cdot$ dispatch / routing};

  \node[mod, below=18mm of cs] (he) {Handoff\\Evaluator};
  \node[mod, left=2mm of he]  (qc) {Quadrant\\Classifier};
  \node[mod, right=2mm of he] (dc) {Deadlock\\Classifier};
  \node[font=\footnotesize\bfseries, above=1.2mm of he] (title)
        {WMARF Mediator \;\scriptsize(augments, not replaces)};
  \begin{scope}[on background layer]
    \node[draw, rounded corners, fill=blue!5, inner sep=2.5mm,
          fit=(qc)(dc)(title)] (wmarf) {};
  \end{scope}

  \node[layer, minimum width=40mm, below=10mm of wmarf] (ale)
    {\textbf{ALE} \scriptsize[existing]\\[-1pt]\scriptsize onboard sensors $\cdot$ local model};
  \node[layer, minimum width=32mm, right=9mm of ale] (tgt)
    {\textbf{Target equipment/zone}\\[-1pt]\scriptsize conveyor / crane / transfer point};

  \draw[arr] ([xshift=-13mm]cs.south) -- ([xshift=-13mm]wmarf.north)
        node[lbl, midway, left, align=right] {state +\\recency};
  \draw[arr] ([xshift=13mm]wmarf.north) -- ([xshift=13mm]cs.south)
        node[lbl, midway, right, align=left] {reroute /\\hold};
  \draw[arr] ([xshift=-13mm]ale.north) -- ([xshift=-13mm]wmarf.south)
        node[lbl, midway, left, align=right] {telemetry,\\sensor};
  \draw[arr] ([xshift=13mm]wmarf.south) -- ([xshift=13mm]ale.north)
        node[lbl, midway, right, align=left] {authority\\token};
  \draw[arr] (ale.east) -- (tgt.west) node[lbl, midway, above] {interact};
\end{tikzpicture}%
}
  \caption{WMARF deployment. The mediator augments the existing CS and decides,
  per interaction, whether the CS or the ALE holds authority; it never moves
  equipment or rewrites control logic.}
  \label{fig:arch}
\end{figure}

WMARF \emph{augments}, rather than replaces, an existing CS: a thin mediator
runs alongside the CS, with a lightweight client on each ALE
(Fig.~\ref{fig:arch}). It reads the CS world-model entry for a target together
with its timestamp, and the ALE's proximity and onboard sensor reading; it
emits an \emph{authority token} -- CS-holds or ALE-holds -- for that interaction,
and, on trouble, a deadlock label and a resolution. The CS keeps dispatching
and the ALE keeps acting (e.g., driving); WMARF only decides whose decision counts, and pushes
a corrected state back when the ALE's sensors win. Integration uses the fleet's
existing interface (e.g., VDA~5050), so no control logic is rewritten
and no shared protocol is imposed.

\subsection{Implementation and scenario}
We implemented the three components of Fig.~\ref{fig:arch} as Python
modules (a Quadrant Classifier, a Handoff Trigger Evaluator, and a Deadlock
Classifier) driven by a discrete-event simulation in SimPy. The scenario is
the canonical hard case: the CS dispatches two ALE to the same
\emph{semi-automated} transfer point, which it believes is \emph{ready} from a
stale update, while the point is in fact still in a mechanical transition that
its interface does not expose. WMARF classifies the situation as Q2, a good
global picture but an under-represented internal state, where the handoff
protocol should engage.

\subsection{Result}
\begin{figure}[t]
  \centering
  \includegraphics[width=\columnwidth]{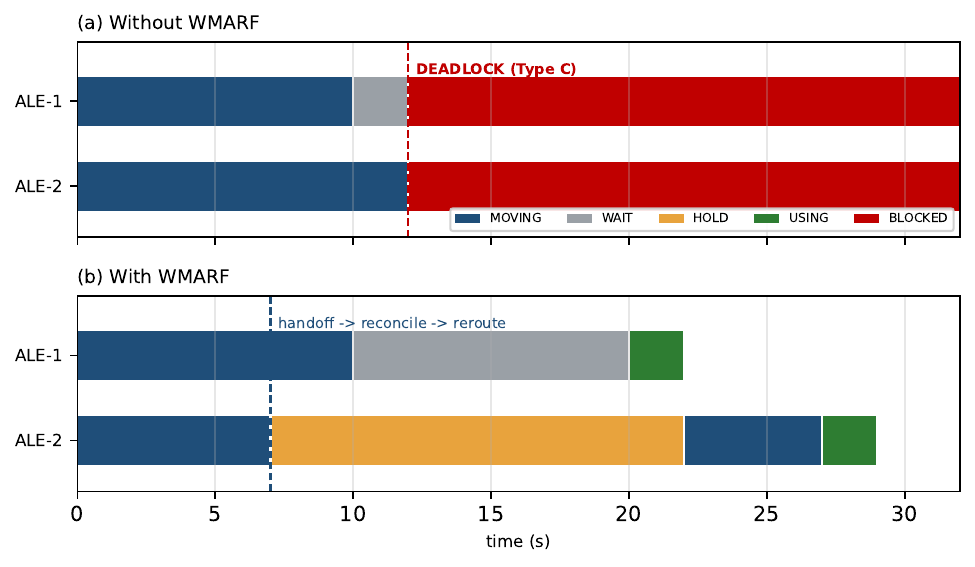}
  \caption{The transfer-point scenario, the same world run twice.
  (a)~Without WMARF, both ALE commit on the stale model and a Type-C divergence
  deadlock occurs at $t=12$\,s. (b)~With WMARF, a proximity handoff at $t=7$\,s
  reconciles the CS model and holds the second ALE, and both transfers
  complete. All values are produced by running the simulation.}
  \label{fig:result}
\end{figure}

Figure~\ref{fig:result} reports both runs of the same world. \emph{Without}
WMARF, both ALE commit on the stale ``ready'' belief; the second commits at
$t=12$\,s while the point is not serviceable, and the Deadlock Classifier
labels a \textbf{Type-C} deadlock (authority held on divergent models), from
which neither transfer completes. \emph{With} WMARF, the first ALE crosses the
proximity threshold at $t=7$\,s, the Handoff Evaluator transfers authority to
it, its sensor reading reconciles the CS model, and the CS holds the second
ALE; both transfers then complete and no deadlock occurs. The same trigger
logic that surfaces the divergence also prevents it -- evidence for RQ1
(dynamic authority averts the failure) and RQ2 (the deadlock is classifiable by
authority state). This is a single, qualitative demonstration; a quantitative
study across scenarios is future work.

\subsection{Reproduction on the VDA~5050 interface}
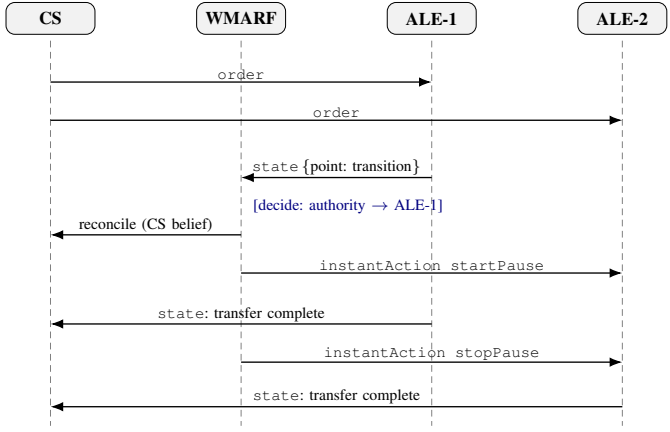
\begin{figure}[t]
  \centering
  \resizebox{\columnwidth}{!}{
\begin{tikzpicture}[
  font=\footnotesize,
  >={Latex[length=2mm]},
  head/.style={draw, rounded corners, fill=black!5, minimum width=14mm,
               minimum height=5mm, font=\footnotesize\bfseries},
  msg/.style={->, semithick},
  lbl/.style={font=\scriptsize, inner sep=1pt},
]
  \def\CS{0} \def\WM{3} \def\AO{6} \def\AT{9} \def\bot{-6.4}
  \node[head] (cs) at (\CS,0) {CS};
  \node[head] (wm) at (\WM,0) {WMARF};
  \node[head] (a1) at (\AO,0) {ALE-1};
  \node[head] (a2) at (\AT,0) {ALE-2};
  \foreach \x in {\CS,\WM,\AO,\AT}
    \draw[densely dashed, black!50] (\x,-0.35) -- (\x,\bot);

  \draw[msg] (\CS,-1.0) -- (\AO,-1.0) node[lbl, midway, above] {\texttt{order}};
  \draw[msg] (\CS,-1.6) -- (\AT,-1.6) node[lbl, midway, above] {\texttt{order}};
  \draw[msg] (\AO,-2.5) -- (\WM,-2.5)
        node[lbl, midway, above] {\texttt{state}\,\{point: transition\}};
  \node[lbl, anchor=west, text=blue!50!black] at (\WM+0.15,-2.95)
        {[decide: authority $\to$ ALE-1]};
  \draw[msg] (\WM,-3.4) -- (\CS,-3.4)
        node[lbl, midway, above] {reconcile (CS belief)};
  \draw[msg] (\WM,-4.0) -- (\AT,-4.0)
        node[lbl, midway, above] {\texttt{instantAction startPause}};
  \draw[msg] (\AO,-4.8) -- (\CS,-4.8)
        node[lbl, midway, above] {\texttt{state}: transfer complete};
  \draw[msg] (\WM,-5.4) -- (\AT,-5.4)
        node[lbl, midway, above] {\texttt{instantAction stopPause}};
  \draw[msg] (\AT,-6.1) -- (\CS,-6.1)
        node[lbl, midway, above] {\texttt{state}: transfer complete};
\end{tikzpicture}}
  \caption{VDA~5050 message exchange for the WMARF reproduction: the CS issues
  \texttt{order}s; ALE-1's \texttt{state} reveals the divergence; WMARF takes
  the handoff, reconciles the CS belief, and sequences ALE-2 with
  \texttt{startPause}/\texttt{stopPause}, all over the standard interface, the
  WMARF core unchanged.}
  \label{fig:vda5050_seq}
\end{figure}

To confirm that WMARF attaches to a real industrial interface and not only to
the simulator, we re-ran the same scenario over \emph{VDA~5050}~\cite{vda5050},
the standard interface between a master control and AGVs
(Fig.~\ref{fig:vda5050_seq}). Only the input/output adapter changes: the
Quadrant Classifier, Handoff Evaluator, and Deadlock Classifier are reused
\emph{unchanged}, so WMARF augments the existing interface rather than
replacing control logic. The ALE \texttt{state} heartbeat is the recency
signal; the bracketed \texttt{[decide:]} annotation in Fig.~\ref{fig:vda5050_seq}
marks WMARF's internal authority decision, which triggers the subsequent
handoff messages that reconcile the CS belief and sequence the second ALE
with \texttt{instantActions}. The run reproduces Fig.~\ref{fig:result}, the
same Q2 handoff prevents the Type-C deadlock, over an in-process broker that
can be swapped for MQTT.

\section{Discussion and Outlook}
WMARF reframes a class of logistics deadlocks as \emph{governance} failures
and shows, on one scenario, that assigning authority by world-model quality
averts a divergence deadlock that static control does not. The evidence is
deliberately limited: a single qualitative scenario with two ALE and one
target, a minimal deadlock model, and a mediator assumed reliable and above the
safety-critical path. Above all, the trigger thresholds $\theta$, $\delta$, and
$\tau$ are set by hand; identifying the full trigger set and calibrating these
thresholds is the principal open problem. The trigger taxonomy of
Table~\ref{tab:triggers} is domain-agnostic; its thresholds are the
adaptation point for each deployment.

Three lines of work follow: a quantitative study of WMARF against static
centralized and decentralized baselines -- deadlock rate, throughput, and
resolution time across heterogeneity levels; a formal treatment of the handoff
protocol, covering convergence, deadlock-freedom, and principled threshold
calibration; and a testbed -- a fleet manager over
VDA~5050~\cite{vda5050} and Open-RMF~\cite{openrmf}, whose resource
locking WMARF complements with an authority layer -- scaled to many ALE and targets
and richer non-autonomous-equipment state. Because
the same thresholds can be retuned to widen the ALE-dominant zone as equipment
matures, WMARF offers an allocation that stays valid as autonomy grows, a key practical advantage.

\bibliographystyle{IEEEtran}
\bibliography{bibliography}

\end{document}